\documentclass[%
 reprint,
superscriptaddress,
nofootinbib,
 amsmath,amssymb,
 aps,
 prd,
floatfix
]{revtex4-2}

\usepackage{hyperref}
\usepackage{xcolor}
\usepackage{ mathrsfs }
\usepackage{graphicx}
\usepackage{dcolumn}
\usepackage{bm}
\usepackage{multirow} 
\usepackage{array} 
\usepackage{cprotect}
\usepackage{dsfont}

\usepackage[commentmarkup=uwave]{changes}
\definechangesauthor[name=MI, color=cyan]{MI}
\definechangesauthor[name=HS, color=orange]{HS}
\definechangesauthor[name=WF, color=magenta]{WF}

\newcommand{\inprod}[2]{\left\langle \mathbf{#1} \vert \mathbf{#2} \right\rangle}

\begin{document}


\title{Analyzing black-hole ringdowns II: data conditioning}

\newcommand{\sbuaffil}{\affiliation{Department of Physics and Astronomy, Stony Brook University, Stony Brook NY 11794, USA}}
\newcommand{\ccaaffil}{\affiliation{Center for Computational Astrophysics, Flatiron Institute, New York NY 10010, USA}}
\newcommand{\cuaffil}{\affiliation{
 Department of Physics, Columbia University,
704 Pupin Hall, 538 West 120th Street, New York, New York 10027, USA
}}
\author{Harrison Siegel}
\email{hs3152@columbia.edu}
\cuaffil{}
\ccaaffil{}
\author{Maximiliano Isi}%
 \email{misi@flatironinstitute.org}
\ccaaffil{}
\author{Will M. Farr}%
 \email{will.farr@stonybrook.edu}
\ccaaffil{}
\sbuaffil{}

\begin{abstract}
Time series data from
observations of black hole ringdown gravitational waves are often analyzed in the time domain by using damped sinusoid models with acyclic boundary conditions. Data conditioning operations, including downsampling, filtering, and the choice of data segment duration, reduce the computational cost of such analyses and can improve numerical stability. Here we analyze simulated damped sinsuoid signals to illustrate how data conditioning operations, if not carefully applied, can undesirably alter the analysis' posterior distributions.  We discuss how currently implemented downsampling and filtering methods, if applied too aggressively, can introduce systematic errors and skew tests of general relativity. These issues arise because current downsampling and filtering methods do not operate identically on the data and model. Alternative downsampling and filtering methods which identically operate on the data and model may be achievable, but we argue that the current operations can still be implemented safely. We also show that our preferred anti-alias filtering technique, which has an instantaneous frequency-domain response at its roll-off frequency, preserves the structure of posterior distributions better than other commonly used filters with transient frequency-domain responses. Lastly, we highlight that exceptionally long data segments may need to be analyzed in cases where thin lines in the noise power spectral density overlap with central signal frequencies. Our findings may be broadly applicable to any analysis of truncated time domain data with acyclic boundary conditions.
\end{abstract}

\maketitle

\section{\label{sec:Intro}Introduction}
In the theory of general relativity, perturbed black holes produce gravitational wave signals consisting primarily of a sum of individual damped sinusoids. This emission is referred to as the \emph{ringdown}, and the individual damped sinusoids are called quasinormal modes (QNMs) \cite{Teukolsky:1973ha, Andersson_BHPerturbationSurvey1999, Teukolsky:Kerrmetric}. Astrophysical binary black hole mergers create perturbed remnant black holes~\cite{Owen:2009sb, Campanelli:2008dv} which subsequently emit ringdown radiation that is observable by detectors like LIGO, Virgo, and KAGRA (LVK)~\cite{AdvancedLIGOScientific:2014pky, VIRGO:2014yos, KAGRA:2020tym}. Intense data analysis efforts have been undertaken to fit the QNMs of black hole ringdown signals in the LVK catalogue \cite{GW150914_detection,Carullo:2019flw,IsiNoHair_GW150914,Isi_BHArea,TestingGR_LIGO_2ndCatalog,Isi_revisitGW150914, Isi:2023nif, Gregorio_ringdown, FinchMoore_GW150914, Ma:2023vvr, CapanoGW190521g_330, YiFan_FreqDomGW150914Overtone,LIGOScientific:2021sio,Siegel:2023lxl}, motivated in part by hopes that these fits can constrain strong field predictions of general relativity. Several time-domain analysis techniques have been devised to fit QNMs~\cite{AnalyzingBHRingdowns, Pyring, Bhagwat_RingdownSBITimeDomain}. Additionally, Ref.~\cite{CapanoGW190521g_330} employs a frequency-domain technique which is formally equivalent to the time-domain analysis of~\cite{AnalyzingBHRingdowns, Pyring}; other frequency-domain methods include those in Refs.~\cite{FinchMoore_FreqDomainAnalysisRingdown, Ma_RationalFilter, Crisostomi_SBIGW150914RingdwonFreqDomain}.

In this paper, we discuss data analysis techniques and challenges associated with time-domain fitting of time series data consisting of exponentially damped sinusoids in Gaussian noise. In particular, we focus on data conditioning (i.e., the operations of downsampling, filtering, and the choice of data segment duration), and the issues which can arise in our ringdown analysis from these operations.  Although we study these effects using the \textsc{ringdown} analysis code \cite{ringdown_code}, our conclusions are broadly applicable, including to analyses of non-ringdown time-domain data, such as in Refs.~\cite{Isi_BHArea,Miller:2023ncs}. The data conditioning issues discussed here are especially relevant for hierarchical analyses of multiple signals, where any untamed biases can accumulate and easily dominate the hierarchical model posteriors~\cite{Isi_HierarchicalTestGR,Moore_InsidiousGRHierarchical}. Controlling such conditioning issues will also be essential for data from the next generation of gravitational wave detectors~\cite{LIGOvoyager, ETDesignReport, CE_Horizon_study, Baker:2019nia}, because these instruments will be an order of magnitude more sensitive than LIGO.

Through analyses of simulated signals, in this paper we demonstrate how the posterior probability distribution can be undesirably altered when insufficiently high sample rates or insufficiently long data segments are used. If standard downsampling and filtering methods are not carefully applied, we find that conditioning-induced posterior alterations can lead to false-positive detections of deviations from general relativity. We also describe our preferred anti-alias filter, dubbed the ``digital filter," which has an instantaneous frequency-domain response at its roll-off frequency. The digital filter is found to be better at preserving posterior structure than traditional analog-like anti-alias filters commonly used in gravitational wave data analysis, such as the Chebyshev or Butterworth filters. Lastly we show that, in order to recover the full signal-to-noise ratio, exceptionally long data segments may need to be analyzed in cases where sharp lines in the noise power spectral density (PSD) overlap with central signal frequencies. We note that removing these sharp lines before performing an analysis can significantly reduce the required data segment length.

Throughout this work, we offer useful mathematical (Eq.~\eqref{eq:logL_differences}) and graphical diagnostics (Sec.~\ref{sec:freq_domain}, Figs.~\ref{fig:PSD_estimation},~\ref{fig:downsampling_noNoise},~\ref{fig:PSD_line},~\ref{fig:cumulativeSNR_line},~and~\ref{fig:appendix_downsampling_noNoise}) to guide analysts and connect time-domain ringdown analysis intuitions with those of more familiar gravitational-wave data analyses, which are typically performed in the Fourier domain. Eq.~\eqref{eq:logL_differences} states that, so long as data conditioning does not change by close to or more than $\mathcal{O}(1)$ the log-likelihood differences of pairs of points over all of parameter space between a conditioned and ``unconditioned'' analysis (i.e., an analysis with the maximum possible sample rate, and an infinitely long data segment), then the posteriors of the conditioned and unconditioned analyses can be considered equivalent. The graphical diagnostics used throughout help to indicate what types of signals are more sensitive to data conditioning, and why.

The paper is organized as follows. In Sec.~\ref{sec:likelihood} we describe our time-domain likelihood and model, and we explain both a statistical criterion for determining safe conditioning settings and a way to approximately visualize our analysis in the frequency domain to gain additional intuition. In Sec.~\ref{sec:DS_Filt} we explore the posterior alterations induced by different downsampling and filtering methods, and we show how a careful application of our currently implemented methods is required to avoid false-positive detections of deviations from general relativity. In Sec.~\ref{sec:PSD_lines} we demonstrate that long data segment durations are needed in order to recover the total signal-to-noise ratio (SNR) whenever there are narrow spectral features (``lines'')  in the PSD near the central frequencies of the signal. In Sec.~\ref{sec:conclusion} we give concluding remarks. In Apps.~\ref{sec:AppendixA}~and~\ref{sec:AppendixB} we illustrate the way in which the posterior alterations caused by downsampling depend sensitively on QNM parameters like phase, and we address a technical point related to a simplification of our model which depends on the number of interferometers used in the analysis.

A data release for this paper can be found in Ref.~\cite{data_release}. Our notational convention in this paper is to denote vectors by lowercase bold letters, and matrices by uppercase bold letters. The components of vectors and matrices are denoted with indices and are not bolded, and retain their letter case.

\section{Time-Domain Likelihood and Signal Model}\label{sec:likelihood}

\subsection{Likelihood}\label{sec:subsection_likelihood}
The log-likelihood function of our Bayesian analysis~\cite{AnalyzingBHRingdowns} is formulated in the time domain as
\begin{align}
\begin{split}
    \text{ln}\, \mathscr{L}\big(\textbf{d} \mid\textbf{s}(\boldsymbol{\psi)}\big) =& -\frac{1}{2} \sum_{i,j = 0}^{N-1} \left(d_i-s_i\right) C_{ij}^{-1} \left(d_j-s_j\right) \\ &+ \text{const.},
    \label{eq:logL}
\end{split}
\end{align}
for a time series data segment $\mathbf{d}$ of length $N$ which is fit with a signal model $\mathbf{s}$ dependent on a set of parameters $\boldsymbol{\psi}$, where $\mathbf{C}$ is the noise covariance matrix which---crucially---will have acyclic boundary conditions in general. The constant in Eq.~\eqref{eq:logL} equals ${-\frac{1}{2} \, \text{ln} \, \text{det} \mathbf{C} -\frac{N}{2} \, \text{ln} \, 2 \pi}$. If the data has noise $\mathbf{n}$, then ${\mathbf{d}=\mathbf{n}+\mathbf{s_T}}$, where $\mathbf{s_T}$ is the true signal, which may or may not be representable by our model $\mathbf{s}$ (although in this paper, our models will always be from the same class of functions as the true signal). For stationary noise, the covariance matrix $\mathbf{C}$ can be derived from an autocovariance function (ACF), which encodes covariance as a function of time lag. The ACF can either be estimated directly in the time domain by taking the sample autocorrelation of a stretch of noise, or alternatively by first estimating the noise PSD and then Fourier transforming since the Wiener-Khinchin theorem states that the ACF and PSD are Fourier transform pairs. 

We now formalize several useful tools which will allow us to re-express the likelihood in an instructive alternate form. We can Cholesky decompose~\cite{Num_recipes} $\mathbf{C}$ by defining a lower triangular matrix $\mathbf{L}$ such that
\begin{equation}
    C_{ij} = \sum_k L_{ik} L_{kj} \, .
    \label{eq:cholesky_decompose}
\end{equation}
The matrix $\mathbf{L^{-1}}$ acts on a given time series $\mathbf{x}$ drawn from a multivariate normal with covariance $\mathbf{C}$ to produce an uncorrelated, i.e. whitened, time series $\mathbf{\bar{x}}$:
\begin{equation}
    \bar{x}_i = \sum_j L_{ij}^{-1} x_j \, ,
    \label{eq:whitened_timeseries}
\end{equation}
whose terms are independent draws from a standard normal distribution. We define a noise-weighted inner product~\cite{guidetoLV_noise} between arbitrary time series $\mathbf{x}$ and $\mathbf{y}$ as follows:
\begin{equation}
    \inprod{x}{y}= \sum_{i,j = 0}^{N-1} x_i C_{ij}^{-1} y_j = \sum_{i = 0}^{N-1} \bar{x}_i \bar{y}_i \, .
    \label{eq:inprod}
\end{equation}
The optimal SNR of a timeseries $\mathbf{x}$ is defined by
\begin{equation}
    \text{SNR}_\text{opt}(\mathbf{x}) = \sqrt{\inprod{x}{x}} \, .
    \label{eq:optimalSNR_def}
\end{equation}
For a noisy timeseries ${\mathbf{d}=\mathbf{n}+\mathbf{s_T}}$, the matched-filter SNR of a signal model $\mathbf{s}$ (not necessarily equal to $\mathbf{s_T}$) is given by
\begin{equation}
    \text{SNR}_\text{mf}(\mathbf{d}, \mathbf{s}) = \frac{\hfill \inprod{s}{d}}{\sqrt{\inprod{s}{s}}} \, .
    \label{eq:matchedfiltSNR_def}
\end{equation}

Using the expressions above, Eq.~\eqref{eq:logL} becomes
\begin{align}
\begin{split}
    \text{ln}\, \mathscr{L}(\textbf{d}|\textbf{s}) = & -\frac{1}{2} \Big(\inprod{d}{d}-2\inprod{d}{s} + \inprod{s}{s}\Big)\, \\ &+ \text{const.}.
    \label{eq:logL2}
\end{split}
\end{align}
Evidently, the three key terms in the log-likelihood are proportional to $\text{SNR}_\text{opt}^2(\mathbf{d})$, ${\text{SNR}_\text{mf}(\mathbf{d}, \mathbf{s})\cdot \text{SNR}_\text{opt}(\mathbf{s})}$, and $\text{SNR}_\text{opt}^2(\mathbf{s})$.

Now let us consider generically the effects of conditioning operations. We are not required to choose the same conditioning operations for the model and data, the only restriction being that the operations for the model and data must both return timeseries with the same number of data points $N$: we denote the matrix of all conditioning operations for the model by $\mathbf{B_s}$, and all those for the data by $\mathbf{B_d}$. This distinction between the two conditioning operations is relevant because many time-domain ringdown analyses~\cite{Pyring, AnalyzingBHRingdowns, CapanoGW190521g_330, IsiNoHair_GW150914, Carullo:2019flw, Siegel:2023lxl} have been performed with ${\mathbf{B_s}\neq\mathbf{B_d}}$, the implications of which we will discuss further in Sec.~\ref{sec:DS_Filt}. The operations $\mathbf{B_s}$ and $\mathbf{B_d}$ act on the unconditioned~\footnote{By unconditioned, we mean no downsampling or filtering and an infinite data segment duration.} vectors $\mathbf{s}$ and $\mathbf{d}$ such that the log-likelihood takes the form of ${\text{ln}\, \mathscr{L}_{\text{cond}} (\textbf{d}|\,\textbf{s}) \equiv \text{ln}\, \mathscr{L}(\mathbf{B_d}\,\textbf{d}|\mathbf{B_s}\,\textbf{s})}$. Since the covariance should match the data, the unconditioned noise covariance matrix $\mathbf{C}$ is also implicitly acted on by the conditioning matrices, such that it enters $\text{ln}\, \mathscr{L}_{\text{cond}}$ as ${\mathbf{C_{\text{cond}}} (\textbf{d}|\,\textbf{s}) \equiv \mathbf{B_d}\, \mathbf{C}\, \mathbf{B_d^T}}$. We will denote the unconditioned log-likelihood by ${\text{ln}\, \mathscr{L}_{\text{unc}}(\textbf{d}|\,\textbf{s}) \equiv \text{ln}\, \mathscr{L}(\textbf{d}|\,\textbf{s})}$.

The goal of this paper is to determine how much downsampling and data segment length truncation can be applied while keeping the structure of the posteriors unchanged. To this end, we note that the $\inprod{d}{d}$ term in Eq.~\eqref{eq:logL2} depends only on the data itself, and as such is a constant offset which does not affect inferences of the parameters $\boldsymbol{\psi}$. Thus, we are interested in how data conditioning might alter the $\inprod{d}{s}$ and $\inprod{s}{s}$ terms in Eq.~\eqref{eq:logL2}. With these considerations in mind, we arrive at an important point: data conditioning should not significantly alter the posteriors of our analysis as long as, over all possible pairs of explored parameters $\boldsymbol{\psi_1}$ and  $\boldsymbol{\psi_2}$,
\begin{align}
\begin{split}
    \biggl| & \text{ln}\mathscr{L}_{\text{cond}}\big(\textbf{d} \mid\textbf{s}(\boldsymbol{\psi_1)}\big)-\text{ln}\mathscr{L}_{\text{unc}}\big(\textbf{d} \mid\textbf{s}(\boldsymbol{\psi_1)}\big) \\ &-\Bigl[\text{ln}\mathscr{L}_{\text{cond}}\big(\textbf{d} \mid\textbf{s}(\boldsymbol{\psi_2)}\big)-\text{ln}\mathscr{L}_{\text{unc}}\big(\textbf{d} \mid\textbf{s}(\boldsymbol{\psi_2)}\big)\Bigr]\biggr| \ll 1
    \label{eq:logL_differences}
\end{split}
\end{align} 
For intuition regarding the above criterion, consider Metropolis-Hastings sampling of the parameter space using the unconditioned and conditioned likelihoods. For Metropolis-Hastings sampling to be unaffected by conditioning, we require ${\mathscr{L}_{\text{unc}}(\boldsymbol{\psi_1}) / \mathscr{L}_{\text{unc}}(\boldsymbol{\psi_2})\sim\mathscr{L}_{\text{cond}}(\boldsymbol{\psi_1}) / \mathscr{L}_{\text{cond}}(\boldsymbol{\psi_2})}$ for all explored parameters ${\boldsymbol{\psi_1}\neq\boldsymbol{\psi_2}}$. This implies Eq.~\ref{eq:logL_differences}.  When the log-likelihood difference is bounded by ${\varepsilon \ll 1}$, the difference in proposal acceptance probability is similarly bounded. Then, sampling with both likelihoods using the same randomness and starting at the same parameters will proceed through the same sequence of parameter samples for, on average, more than $N = 1/\varepsilon$ steps. By contrast, $\mathcal{O}(1)$ log-likelihood differences lead to divergence in the first few steps.  Over a large number of steps, the chains become well-mixed in any case, so a single trajectory deviation does not matter; but a deviation after only a few steps can lead to wildly divergent, statistically inequivalent samplings.

In practice, computing Eq.~\ref{eq:logL_differences} exactly is difficult without running the full analysis. One option for estimating log-likelihood changes before performing the full analysis is to use the ringdown portion of an IMR analysis as a proxy for $\mathbf{s}$, then track the log-likelihood changes induced when that proxy signal is used in downsampling, filtering, and data segment length truncation, and determine from that some possible ``safe'' conditioning settings which will not alter the posteriors in the real analysis. Once a ringdown analysis is completed with these initial settings, the log-likelihood of that analysis can then be re-computed with more conservative conditioning to see if this would have led to a significant change.

\subsubsection{No-noise analyses}\label{sec:no-noise-anal}

In this paper, we will restrict ourselves to analyzing no-noise data such that ${\mathbf{n}=\mathbf{0}}$ and $\mathbf{C}$ sets the SNR. We focus on no-noise analyses for demonstrative clarity, as they make it more straightforward to assess the systematic effects of our conditioning operations. In a no-noise injection, as long as 1) the correct model for the data is used; 2) the prior is flat around the truth; and 3) the data conditioning does not alter the posteriors at all; then the maximum likelihood point of our full multidimensional posterior should coincide with the truth. Marginalization can make lower-dimensional low-SNR distributions not peak at the truth, but we are generally looking at high SNR analyses in this paper. With a specific noise instantiation included in the data, the maximum likelihood point will typically not coincide with the true value even if conditioning is not altering the posteriors. The no-noise analyses essentially give the expectation over many different noise realizations. For readers interested in performing injections with noise, in the data release for this paper~\cite{data_release} we provide methods for generating simulated time series with stationary noise drawn from any prescribed power spectral density.

When analyzing no-noise data, it is sufficient to just check for differences in the $\inprod{s}{s}$ term of Eq.~\eqref{eq:logL2} in order to evaluate Eq.~\eqref{eq:logL_differences}, i.e., we can just compute optimal SNR changes. In general however, when there is noise, such that ${\mathbf{n}\neq\mathbf{0}}$, it is necessary to evaluate all terms of Eq.~\eqref{eq:logL_differences}, as it is possible for the specific noise instantiation to strongly affect how sensitive the second term in Eq.~\eqref{eq:logL2} is to conditioning.

\subsection{Signal model and review of QNMs}

For completeness, we now briefly review QNMs, and then describe our signal model for fitting them. A reader familiar with this topic should skip ahead to Sec.~\ref{sec:freq_domain}.

A QNM is an exponentially-damped sinusoidal gravitational wave with a spheroidal harmonic angular emission pattern, produced by perturbed black  hole spacetimes. At linear order in perturbation theory and following the labeling in \cite{AnalyzingBHRingdowns}, every possible QNM can be uniquely identified by four values $(p,\ell,m,n)\equiv j$, such that: ${p=+}$ and ${p=-}$ respectively denote prograde and retrograde rotation senses of wavefronts relative to the black hole spin when ${m\neq0}$, and they denote polarization degrees of freedom when ${m = 0}$; $\ell$ and $m$ denote angular content; and $n$ denotes radial content. A QNM with ${n=0}$ is called a fundamental mode, and QNMs with ${n>0}$ are called overtones. Increasing $n$ corresponds to decreasing lifetime (except in the extremal spin limit, where zero-damping modes occur~\cite{Yang:2012pj}). For a given $j$, the QNM has four parameters: an amplitude, a phase, an oscillation frequency and a decay rate, all denoted respectively by ($A_j,\, \phi_j,\, f_j,\, \gamma_j$). All QNM frequencies $f_{j}$ and damping rates $\gamma_{j}$ (or equivalently, damping times ${\tau_{j}\equiv 1/\gamma_{j}}$) are determined in general relativity solely by the mass and spin $(M,\chi)$ of the background Kerr metric, and can be calculated theoretically to effectively arbitrary precision~\cite{qnmpackage_Stein, Leaver:1985ax, Cook:2014cta}. Each QNM amplitude $A_j$ and phase $\phi_j$ is nontrivially related to the initial conditions of the system~\cite{Zhu:2023fnf, London_ModelingRingdownII, BorhanianRingdown, London_modelingringdownbeyondfundqnms, Kamaretsos_BHHairLoss, Kamaretsos:2012bs, Leaver:1986gd, berti:2006kk}. 

In the rest of this section, we define ${\pm j \equiv (p,\ell,\pm |m|,n)}$ for ${m\neq 0}$, and ${\pm j \equiv (\pm ,\ell,0,n)}$ for ${m=0}$. This choice is motivated by the fact that, for Kerr black holes~\cite{Kerr:Kerrmetric, Teukolsky:Kerrmetric}, given a set of ($p$, $l$, $n$) and ${|m|>0}$, the $\pm m$ modes share an identical $f$ and $\gamma$; for the special case of ${m=0}$, for a given ($l$, $n$), the ${p=\pm}$ modes share identical $f$ and $\gamma$.\footnote{In general relativity, the complex Kerr QNM frequencies ${\tilde{\omega}_j \equiv 2\pi f_j- i\gamma_j}$ satisfy ${\tilde{\omega}_{j}=-\tilde{\omega}_{-j}^*}$. Modified gravity theories may break this condition~\cite{Li:2023ulk}.} In the injection studies throughout this paper, we impose $|m|>0$ and ${p=+}$, subsequently dropping the $p$ index.

To begin constructing our signal model $\mathbf{s}$ as in \cite{AnalyzingBHRingdowns}, we can express the two $\pm j$ QNM polarization degrees of freedom as functions of time $\big(\mathbf{h}^+_{|j|}(t),~ \mathbf{h}^\times_{|j|}(t)\big)$ in the linear polarization basis~\cite{Isi:2022mbx}, such that the full complex-valued gravitational strain is ${\mathbf{h}_{|j|}(t) = \mathbf{h}^+_{|j|}(t) - i\, \mathbf{h}^\times_{|j|}(t)}$. Here we use a subscript $|j|$ to denote any quantity which combines information from the $\pm j$ polarizations, as shown explicitly in Eqs.~\eqref{eq:h+_hx_def2}. The polarizations, when observed at a point on the angular 2-sphere around the source (i.e., as observed from Earth), have the functional form
\begin{subequations}
\begin{align}
    \mathbf{h}^+_{|j|} = A_{|j|}\exp(-\gamma_{j}\mathbf{t}) &\left[\cos(2\pi f_{j} \mathbf{t} + \phi_{|j|})\cos \theta_{|j|} \right. \nonumber \\
    &\left.-\, \epsilon_{|j|} \sin(2\pi f_{j} \mathbf{t} + \phi_{|j|}) \sin \theta_{|j|}\right], 
\end{align}
\begin{align}
    \mathbf{h}^\times_{|j|} =A_{|j|} \exp(-\gamma_{j}\mathbf{t}) &\left[\cos(2\pi f_{j} \mathbf{t} + \phi_{|j|})\sin\theta_{|j|} \right. \nonumber \\
    &\left.+\, \epsilon_{|j|} \sin(2\pi f_{j} \mathbf{t} + \phi_{|j|})\cos \theta_{|j|}\right],
\end{align}
\label{eq:h+_hx_def}
\end{subequations}
where
\begin{subequations}    \label{eq:h+_hx_def2}
\begin{align}
    A_{|j|} &= |A_j|+ |A_{-j}|\, ,\\
    \phi_{|j|} &= \left(\phi_{j}- \phi_{-j}\right)/2\, ,\\
    \theta_{|j|} &= -\left(\phi_{j}+ \phi_{-j}\right)/2\, ,\\
    \epsilon_{|j|} &= \frac{|A_{j}|- |A_{-j}|}{|A_{j}|+ |A_{-j}|}\, .
\end{align}
\end{subequations}
We evaluate the time $t$ in a given detector at evenly spaced points dictated by the sample rate and duration of the data-segment $\mathbf{d}$ in Eq.~\eqref{eq:logL2}. The amplitude $A_{|j|}$ in Eqs.~\eqref{eq:h+_hx_def} and \eqref{eq:h+_hx_def2} has implicitly absorbed the spheroidal harmonic angular dependence of the QNMs.

To finish constructing our signal model, we must encode the detector's orientation with respect to the source, by projecting the polarizations onto the detector through the antenna patterns $F_{+/\times}$ which are functions of the sky location. Our ringdown model for a single interferometer $I$ is thus a sum of $Q$ modes:
\begin{equation}
    \mathbf{s}_I = \sum_{i=0}^{Q-1} \big[F^I_{+}\mathbf{h}^+_{|j_i|} + F^I_{\times}\mathbf{h}^\times_{|j_i|}\big].
    \label{eq:signal_model}
\end{equation}
The posterior for a multi-interferometer analysis is given by Bayes' rule and is proportional to the multiplication of single-interferometer likelihoods under the reasonable assumption that the noise in each interferometer is statistically independent,
\begin{equation}
    p(\boldsymbol{\psi}|\textbf{d}) \propto p(\boldsymbol{\psi})\prod_I \mathscr{L}\big(\textbf{d}_I|\textbf{s}_I(\boldsymbol{\psi})\big).
\end{equation}
Offsets in the time $t$ of each $\mathbf{d}_I$, related to the time taken for waves to propagate between each interferometer, are determined by the sky location. 

We can further build on the model of Eqs.~(\eqref{eq:h+_hx_def}--\eqref{eq:signal_model}) in order to theory-agnostically test general relativity (TGR), by introducing deviation parameters $(\delta f,~\delta\gamma)$~\cite{AnalyzingBHRingdowns, testgr_ringdownbayesian} such that
\begin{align}
\begin{split}
    f_{\text{TGR}} & = f_{\text{Kerr}}\exp(\delta f), \\
    \gamma_{\text{TGR}} & = \gamma_{\text{Kerr}}\exp(\delta \gamma).
    \label{eq:TGR_fgamma}
\end{split}
\end{align}
For beyond-Kerr signals, $\delta f$ and/or $\delta\gamma$ may be non-zero; however, note that non-zero values can also be achieved for Kerr signals through model misspecification, and conversely it is not clear that all beyond-Kerr signal morphologies are guaranteed to be captured by this theory-agnostic parameterization. This TGR model can be used when fitting 2 or more QNMs. Between 1 and $Q-1$ of the $(f_{j},\gamma_{j})$ in the model can be given the freedom to deviate from general relativity without introducing degeneracies.

Our models ignore other possible ringdown signal content, such as the early-time prompt response which appears to be quickly dominated by QNMs in binary black hole mergers, the late-time tail which seems to be very weak relative to the early-time signal, and higher-order nonlinear QNMs which also generally appear to be subdominant~\cite{Andersson_EvolvingTestFieldsInBHGeometry, Frolov_BlackHolePhysicsTextbook, Giesler:2019uxc, Giesler:2024hcr, Okounkova:2020vwu, Mitman:2022qdl, Cheung:2022rbm}. While in principle most of these effects can be added to our models, highly accurate fits to data at current LIGO sensitivities can generally be obtained with linear QNMs alone.

As a technical point, note that although Eq.~\eqref{eq:signal_model} can be further simplified, since any sum of sines and cosines with identical frequencies but arbitrary amplitudes and phases can always be re-expressed as a single sinusoid~\cite{mathworld_harmonicadditiontheorem}, it is preferable to draw posterior samples using Eq.~\eqref{eq:signal_model} when there are multiple detectors. See App.~\ref{sec:AppendixB} for further discussion.

For further details regarding our analysis, see Ref.~\cite{AnalyzingBHRingdowns}.

\subsection{Relationship to frequency domain}\label{sec:freq_domain}
\begin{figure}
    \centering
    \includegraphics[width=0.5\textwidth]{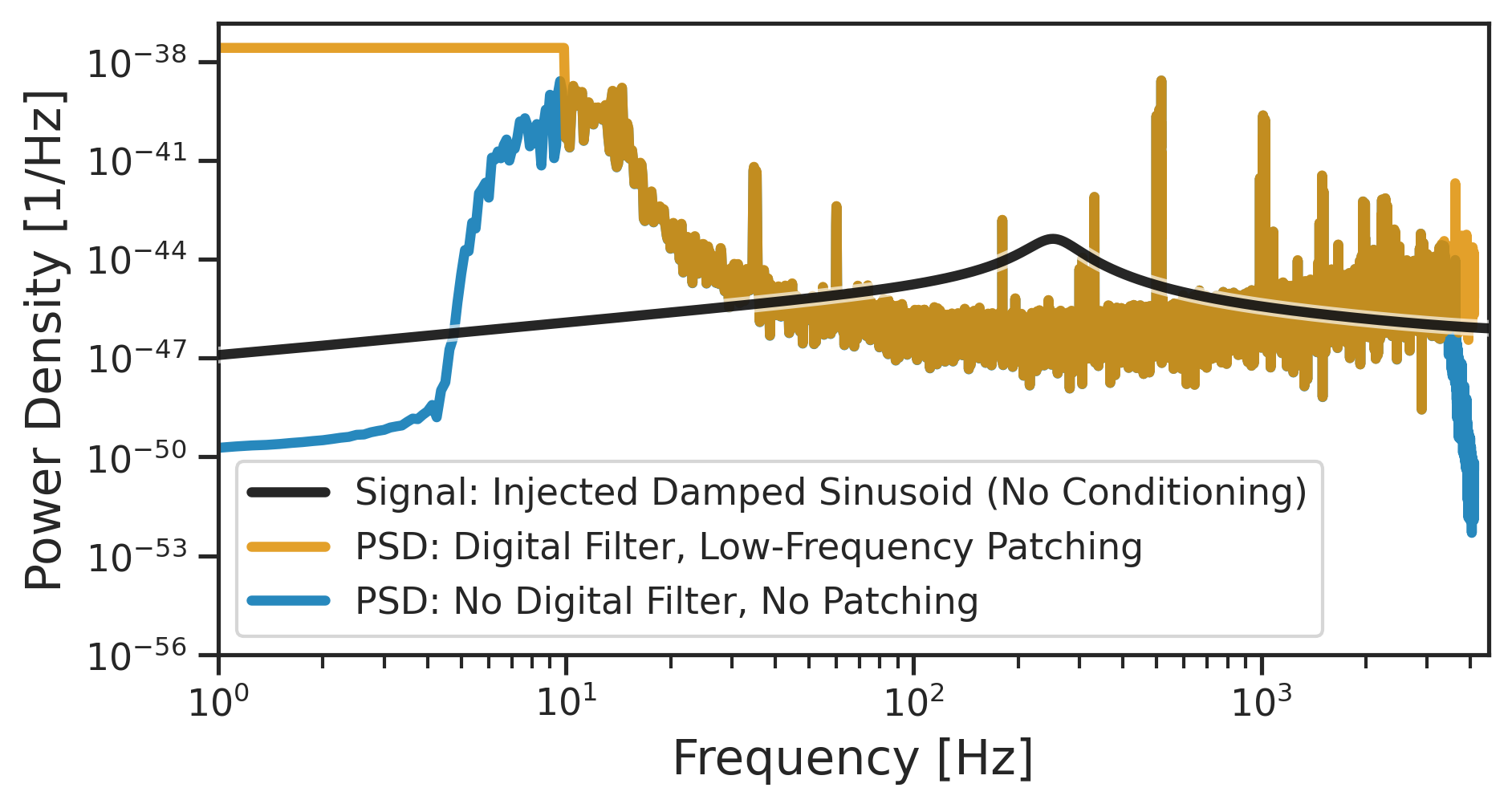}
    \caption{Here we highlight key conditioning issues, and relate them to different Welch estimates of the noise PSD from downsampled GW150914 LVK data~\cite{GW150914_detection} (yellow and blue). We compare these PSDs to a GW150914-like signal~$\mathbf{s}$, shown in black as $f | \tilde{\mathbf{s}}  |^2$ (Sec.~\ref{sec:freq_domain}). As shown in blue, low and high-pass filters with transient frequency-domain responses create dips in the power of low and high-frequency data. These filters impact our posteriors more than filters with instantaneous frequency-domain responses (Sec.~\ref{sec:DS_Filt}). In yellow, we show our preferred way of simultaneously downsampling and low-pass filtering with a method dubbed the ``digital filter,"  and also manually ``patching" the PSD by inflating it at low frequencies. The digital filter applies a tophat function to the data in the frequency domain from 0 to the Nyquist frequency, and Fourier transforms only data below Nyquist. Patching alleviates filtering-induced posterior changes by censoring corrupted low-frequency data in our likelihood. Additionally, thin PSD lines near the signal peak can lower SNR if insufficiently long data segments are analyzed (Sec.~\ref{sec:PSD_lines}). It should be possible to subtract lines from the data due to their long coherence times, allowing for analyses of shorter data segments which still recover the full SNR. }
    \label{fig:PSD_estimation}
\end{figure}

Although our analysis is explicitly formulated in the time domain, it is still possible to build useful intuitions by approximating our likelihood in the Fourier domain. To determine this approximate frequency-domain representation, we manipulate Eq.~\eqref{eq:logL} and make use of the Fourier transform $\Omega_{ij} x_{j} = \tilde{x}_i$ and its unitarity, i.e., $\mathbf{\Omega}^\dagger\mathbf{\Omega} =  \mathds{1} $ where $\mathds{1}$ is the identity matrix. Concretely,
\begin{align}
        \text{ln}\, \mathscr{L}(\mathbf{d}|\mathbf{s}) &= -\frac{1}{2} \left(\mathbf{d}-\mathbf{s}\right)^\dagger \mathbf{C}^{-1} \left(\mathbf{d}-\mathbf{s}\right) \nonumber\\ 
        &= -\frac{1}{2} \left[\left(\mathbf{\Omega}^\dagger\mathbf{\Omega}\right)\left(\mathbf{d}-\mathbf{s}\right)\right]^\dagger\mathbf{C}^{-1} \left[\left(\mathbf{\Omega}^\dagger\mathbf{\Omega}\right)\left(\mathbf{d}-\mathbf{s}\right)\right] \nonumber\\
        &= -\frac{1}{2} \big(\tilde{\mathbf{d}}-\tilde{\mathbf{s}}\big)^\dagger \left(\mathbf{\Omega}\, \mathbf{C} \, \mathbf{\Omega}^\dagger\right)^{-1} \big(\tilde{\mathbf{d}}-\tilde{\mathbf{s}}\big).
    \label{eq:Fourier_domain_logL}
\end{align}
The Fourier transform of each QNM in the signal, $\tilde{\mathbf{s}}$, is well-approximated by a Lorentzian under a discrete Fourier transform. As for the noise covariance matrix, the transformation applied in the last line of Eq.~\eqref{eq:Fourier_domain_logL} tells us that the covariance matrix of the Fourier-domain residuals is $\tilde{\mathbf{C}} = \mathbf{\Omega}\, \mathbf{C}\, \mathbf{\Omega}^\dagger$. In typical LVK analyses, the assumption of stationary noise and, crucially, the enforcement of cyclic boundary conditions, are meant to ensure that this matrix can be treated as diagonal and its entries are proportional to the noise PSD; however, in our case, the lack of cyclic boundary conditions in $\mathbf{C}$ means that $\tilde{\mathbf{C}}$ cannot be diagonal (see also e.g. Ref.~\cite{Talbot:2021igi}). Nonetheless, we expect (and have checked in the case of simulated noise) that diagonal terms in $\tilde{\mathbf{C}}$ dominate and are nearly proportional to the noise PSD, suggesting that we can approximately visualize our analysis in the frequency domain with Lorentzians plotted on top of the noise PSD, even if their ratio does not exactly correspond to our likelihood (as it would in traditional LVK analyses). Specifically, in the continuous-frequency limit, the likelihood approximately becomes an integral over frequency of $| \tilde{\mathbf{s}} |^2 / \ \text{PSD}(f)$, or an integral over log-frequency of $f | \tilde{\mathbf{s}}  |^2 / \ \text{PSD}(f)$; in the plots throughout our paper, we show the latter so that such integrals can be estimated visually by the difference of curves in log space. Figs.~\ref{fig:PSD_estimation},~\ref{fig:downsampling_noNoise},~\ref{fig:PSD_line},~and~\ref{fig:appendix_downsampling_noNoise} depict this approximation, and give valuable insights into how the analysis behaves.

\section{Downsampling and Filtering}\label{sec:DS_Filt}

In this section, we describe the downsampling and filtering procedures that we use in analyses of LVK data; we then implement these techniques to perform studies of simulated signals in the absence of noise. These so-called ``no-noise injections'' still presume a noise covariance matrix $\mathbf{C}$ to set the expected noise level in the likelihood, but impose ${\mathbf{n} = \mathbf{0}}$ so that $\mathbf{d} = \mathbf{s}$ in Eq.~\eqref{eq:logL}; see Sec.~\ref{sec:no-noise-anal} for more details.

The ringdown signals we are most often interested in have central frequencies below 1~kHz, as well as low- and high-frequency tails; see Fig.~\ref{fig:PSD_estimation}, as well as discussion in Sec.~\ref{sec:freq_domain} regarding the frequency domain representation we use (and its limitations). The native sampling rate of calibrated LVK data~\cite{GWOSC} is 16384 Hz; for most ringdown signals, there is seemingly little to gain by utilizing this full frequency range and incurring the associated computational cost, since a relatively small fraction of the total signal power lies at high frequencies. Also, the LIGO detector is less accurately calibrated below 10--20 Hz and above a few kHz (depending on the observing run in question)~\cite{O1O2_Calibration, O3a_Calibration, O3b_Calibration}, meaning that very high and very low frequencies should be viewed with additional caution. Downsampling, and low- and high-pass filtering, are implemented in our analysis to reduce computational cost, avoid analyzing less well-calibrated or overly noisy frequency ranges in our data, and mitigate aliasing.

However, overly aggressive application of downsampling and filtering can lead to SNR loss, i.e., an increase in statistical uncertainty. Additionally, if the downsampling and filtering operations for the signal and data are not the same (as has been the case in many time-domain ringdown analyses~\cite{Pyring, AnalyzingBHRingdowns, CapanoGW190521g_330, IsiNoHair_GW150914, Carullo:2019flw, Siegel:2023lxl}), then the differences between both operations effectively introduce extra terms into the log-likelihood comparison of Eq.~\eqref{eq:logL_differences} which all must be collectively much smaller than 1 in order to not affect the posteriors. This can be most easily seen as follows. We denote the matrix of all conditioning operations for the model by $\mathbf{B_s}$, and all those for the data by $\mathbf{B_d}$. We also denote the difference between both operations as ${\mathbf{B_\xi}\equiv\mathbf{B_s}-\mathbf{B_d}}$. When applying data conditioning, the log-likelihood of Eq.~\eqref{eq:logL2} is modified to give the conditioned log-likelihood, $\text{ln}\, \mathscr{L}_{\text{cond}}$, as
\begin{align}
\begin{split}
    \text{ln}\, \mathscr{L}_{\text{cond}} &\equiv \text{ln}\, \mathscr{L}(\mathbf{B_d}\,\textbf{d}|\mathbf{B_s}\,\textbf{s})\\
     &= \text{ln}\, \mathscr{L}(\mathbf{B_d}\,\textbf{d}|\mathbf{B_d}\,\textbf{s})
     \\ &~~~ + \inprod{B_d d}{B_\xi s} - \inprod{B_d s}{B_\xi s} -\frac{1}{2} \inprod{B_\xi s}{B_\xi s}.
     \label{eq:bias_terms_logL}
\end{split}
\end{align}
Remember that the unconditioned noise covariance matrix $\mathbf{C}$ is also implicitly acted on by the conditioning matrices, such that it enters $\text{ln}\, \mathscr{L}_{\text{cond}}$ as ${\mathbf{C_{\text{cond}}} \equiv \mathbf{B_d}\, \mathbf{C}\, \mathbf{B_d^T}}$. It may be that in some situations, a judicious choice of ${\mathbf{B_\xi}\neq \mathbf{0}}$ could actually be advantageous; well-designed biased estimators can desirably reduce variance~\cite{Efron_Hastie_2021}. However, for the purposes of our analysis, we want to keep our parameter estimation as close as possible to what we would have had without any data conditioning, and thus any manipulation of the variance is not in keeping with our aims. We posit that any information lost from the data through the operation of $\mathbf{B_d}$ in Eq.~\eqref{eq:bias_terms_logL} cannot be recovered for all possible parameters $\boldsymbol{\psi}$ with any choice of $\mathbf{B_\xi}$, and thus the additional terms in Eq.~\eqref{eq:bias_terms_logL} can only serve to increase log-likelihood differences induced by conditioning. Thus, if our conditioned analysis aims to fully preserve the unconditioned likelihood by satisfying Eq.~\eqref{eq:logL_differences}, the additional terms introduced in Eq.~\eqref{eq:bias_terms_logL} when ${\mathbf{B_\xi}\neq \mathbf{0}}$ can only serve to necessitate more conservative conditioning settings (e.g., higher sample rates).

Many previous ringdown analyses~\cite{Pyring, AnalyzingBHRingdowns, CapanoGW190521g_330, IsiNoHair_GW150914, Carullo:2019flw, Siegel:2023lxl} used conditioning operations ${\mathbf{B_d}\neq\mathbf{B_s}}$ because such operations were more straightforward and computationally efficient to implement. One could in principle write a downsampling and filtering method which guarantees that the same operations are applied to the model and whatever data is provided by the LVK~\footnote{Technically, even if the conditioning operations for the data and model are made to be identical, there are still other differences between our data and model which we do not account for, owing to the fact that our signal model does not replicate the entire detector response process through which our data is obtained.}, although the implementation and testing of such a method is beyond the scope of this work. With such alternative methods, the only changes to the posterior when downsampling and filtering would occur when such operations caused signal information in the data to be thrown out. Nonetheless, there are useful regimes in which downsampling and filtering as they are currently implemented should be relatively safe, as we will discuss below. As such, we will proceed to describe our current downsampling and filtering methods.

\subsection{Methods: downsampling and filtering}
For our signal model (Eqs.~\eqref{eq:h+_hx_def}~--~\eqref{eq:signal_model}), we downsample by first evaluating the model at the native sample rate of the data and then simply taking every $n^{\text{th}}$ sample in the time domain (always retaining the first sample of the original time series), where $n$ is the factor by which we are downsampling. This strategy would be a poor choice for the data due to the aliasing of high frequency noise, since the high frequency noise in LIGO data is orders of magnitude larger than the noise at relevant signal frequencies. Thus, a different downsampling procedure is required for the data.

For the data, we simultaneously downsample and perform anti-alias filtering, through a method dubbed the ``digital filter"~\cite{Num_recipes}. The digital filter Fourier transforms the data and applies a top-hat function extending from zero to the target Nyquist frequency, and then Fourier transforms back to the time domain using only frequencies below the Nyquist frequency. We also roll the raw data so that the first time stamp of each interferometer's time series is unchanged after downsampling. This preserves time delays between interferometers. The digital filter contrasts with traditional LVK approaches to downsampling and anti-aliasing, which often involve the use of an analog-like frequency-domain low-pass filter such as the Butterworth or Chebyshev filter. Such analog-like functions generally do not have an instantaneous frequency-domain response at their roll-off frequency; if one were to select the Nyquist frequency as the roll-off frequency when using these analog-like filters, frequencies below the Nyquist frequency would be affected. As a result, noise PSD estimates and actual signal in data filtered with such analog-like functions will have a dip in power near the roll-off frequency (Fig.~\ref{fig:PSD_estimation}). These dips can lead to avoidable and undesirable posterior corruption. Because our digital filtering method better preserves the morphology of the data up to the Nyquist frequency, it is a more stable choice for our analysis.

Currently, publicly available LVK strain data~\cite{GWOSC} has been filtered below ${\sim} 10$~Hz. Internal LVK data may not always have such treatment, in which case we apply the filtering ourselves. High-pass filtering reduces the dynamic range of the PSD, which can increase the numerical stability of the time-domain likelihood. The filtering also removes less well-calibrated low-frequency data.

We currently cannot apply to our signal model the filtering methods we use on our data. This is because the filtering we currently apply to the data convolves the segment targeted for analysis with data points preceding and following it. However, the domain of our signal model time series is limited to only include times targeted for analysis. Thus there are no preceding or following points that the signal model can be convolved with, and we don't apply filtering to the signal model as a result. This inequivalence of conditioning operations between data and model can lead to undesirable posterior alterations, especially due to low frequency data which is more substantially changed by filtering. To get around this issue, we ``patch" our PSD estimate to suppress the low-frequency data: we manually set the PSD to be larger than any affected low-frequency signal power (Fig.~\ref{fig:PSD_estimation}), which causes this data to contribute less significantly in our likelihood. The outcome of this patching is similar to increasing the lower limit of integration in the traditional Fourier-domain likelihood \cite{guidetoLV_noise}. In this paper, our patching routine sets the PSD to be 10 times higher than its maximum value in the patched frequency range. In principle, one may be able to design a filtering operation which only acts on time series points within the targeted analysis segment, and thus can be applied identically to the data and model. We leave this for future work.

\subsection{Injection studies: downsampling and filtering}\label{sec:injectionstudies_downsample}

\begin{figure*}
    \centering
    \includegraphics[width=\textwidth]{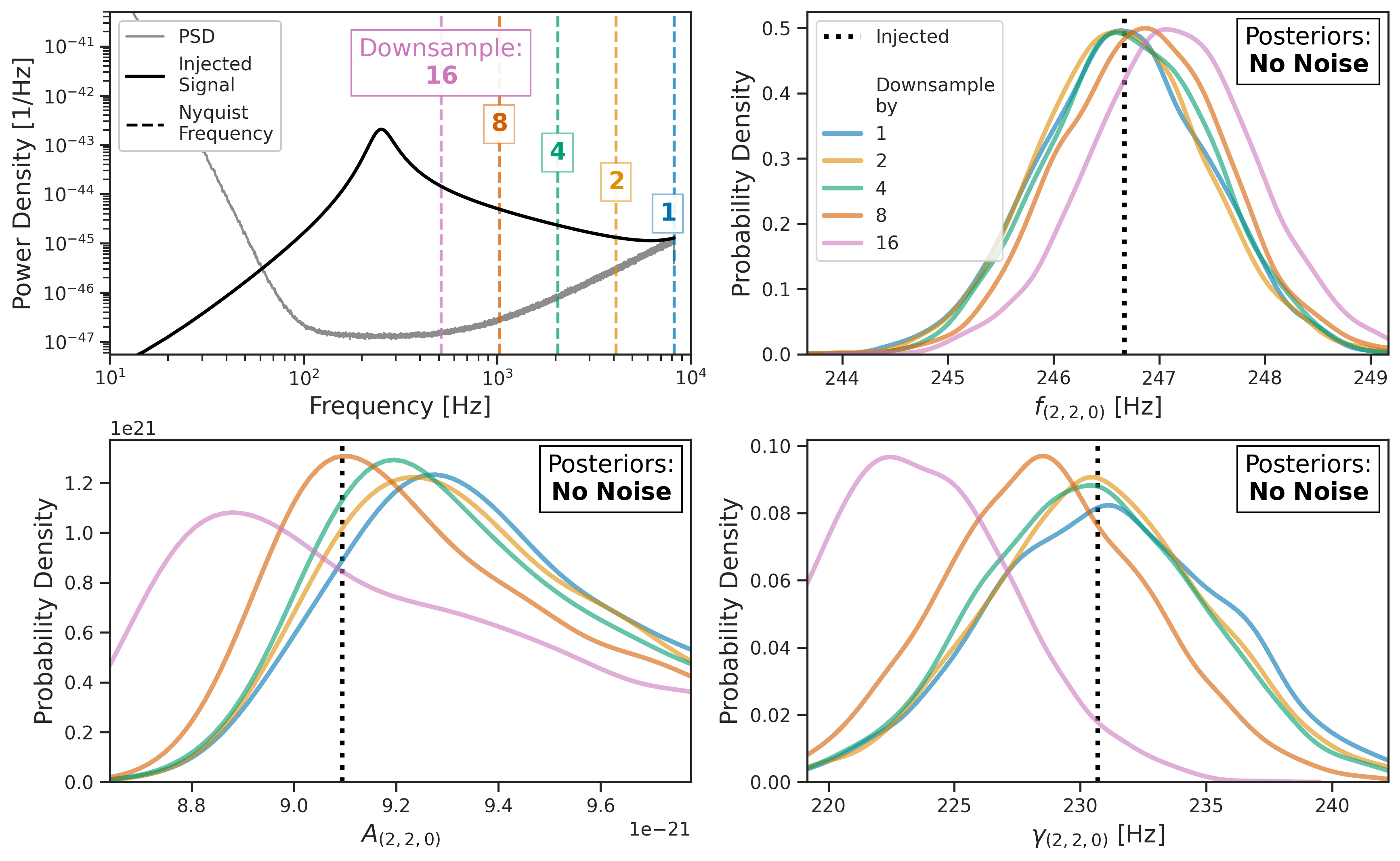}
    \caption{No-noise injections, showing posterior changes which are induced by our currently implemented downsampling method. \textit{Top Left:} We inject a single damped sinusoid with zero noise, and use the aLIGO design PSD~\cite{AdvancedLIGOScientific:2014pky, designPSD_1, designPSD_2} to calculate the noise covariance matrix $\mathbf{C}$ in our likelihood, Eq.~\eqref{eq:logL}. The PSD is shown in grey, the signal $f|\tilde{\mathbf{s}}|^2$ in black, different Nyquist frequencies from downsampling are dashed lines. The injected signal has optimal network ${\text{SNR}\sim 113}$, and parameters like those of the fundamental prograde $(\ell,m)=(2,2)$ QNM of GW150914~\cite{GW150914_detection}; see Table~\ref{tab:injected_params}. \textit{Top Right, Bottom:} Posteriors for QNM parameters, recovered with different downsampling factors using the digital filter. The posteriors change significantly as the downsampling factor increases. These changes depend sensitively on the true QNM parameters; see App.~\ref{sec:AppendixA} for further details. Note that the amplitude posterior does not peak exactly at truth when no downsampling is applied; the amplitude prior is not fully flat around the true value.}
    \label{fig:downsampling_noNoise}
\end{figure*}

\begin{figure*}
    \centering
    \includegraphics[width=\textwidth]{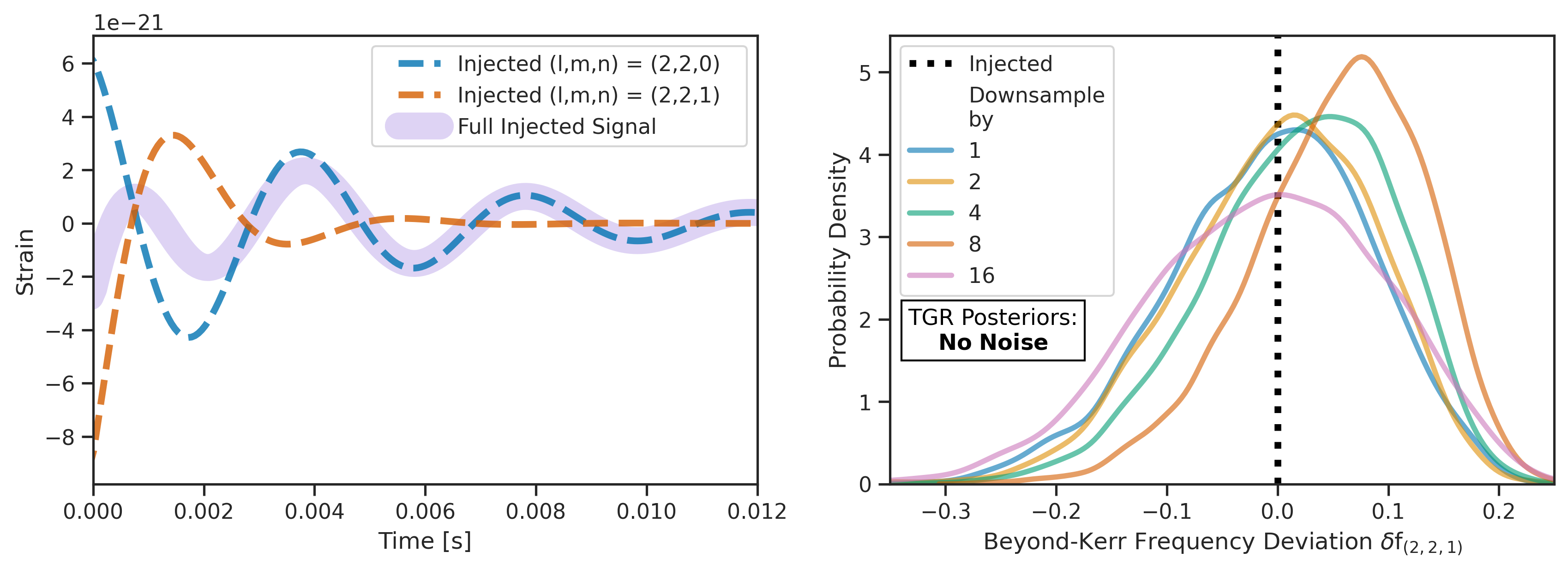}
\caption{Downsampling can cause undesirable posterior changes in our analysis that could be mistaken for deviations from general relativity. \textit{Left:}~We inject two Kerr-consistent GW150914-like QNMs with optimal network ${\text{SNR}\sim57}$, parameters in Table~\ref{tab:injected_params}. The sum of both QNMs is the analyzed signal, shown as a purple curve. Unwhitened strain shown.  \textit{Right:}~No-noise TGR analysis as in Eq.~\eqref{eq:TGR_fgamma}, allowing beyond-Kerr deviations of the $(\ell,m,n) = (2,2,1)$ QNM frequency and damping rate. The noise PSD used to calculate the covariance matrix $\mathbf{C}$ which enters our likelihood, Eq.~\eqref{eq:logL}, is the same as in Fig.~\ref{fig:downsampling_noNoise}, and we downsample with the digital filter. For moderate downsampling factors, the beyond-Kerr frequency deviation posteriors begin peaking away from truth. Such undesirable effects of conditioning may become especially significant in hierarchical analyses.}
    \label{fig:TGR_downsampling}
\end{figure*}

To better understand the posterior changes introduced by our downsampling and filtering methods, we report here on several no-noise injection studies of damped sinusoids. We first investigate single damped sinusoid injections which are downsampled using the digital filter, as shown in Fig.~\ref{fig:downsampling_noNoise}. We then study multi-mode injections which are downsampled with the digital filter and we perform a mock test of general relativity with them, as shown in Fig.~\ref{fig:TGR_downsampling}. Finally, we compare the digital filter with more traditional anti-aliasing filters in Fig.~\ref{fig:digital_vs_analog}.

Our injected signals start with a ``ring-up", which is made of exponentially growing sinusoids, followed immediately by a (phase coherent) ringdown of exponentially decaying sinusoids with the same frequencies as in the ring-up.  The signals are injected into two different simulated interferometers (Hanford and Livingston) from one sky location, and the PSDs of both interferometers are assumed to be identical and are given by the Advanced LIGO (aLIGO) design~\cite{AdvancedLIGOScientific:2014pky, designPSD_1, designPSD_2} as shown in Fig.~\ref{fig:downsampling_noNoise}. Note that this PSD has no lines, unlike the actual LIGO PSD in Fig.~\ref{fig:PSD_estimation}. The start time of our analysis is chosen to be the peak of the signal. The native sample rate of the simulated data is 16384~Hz. We apply anti-alias filtering when downsampling, as well as high-pass filtering and low-frequency PSD patching. For computational efficiency, in this section we analyze data segments of length ${T=0.05}$~seconds, meaning they have at most 820 samples. This is a sufficiently long data segment for the purposes of this specific analysis, as the whitened waveforms accumulate their total power significantly before the end of the data segment (see Fig.~\ref{fig:PSD_line}). For the noise covariance matrix which enters our likelihood~(Eq.~\eqref{eq:logL2}), we need an estimate of the ACF of a long stretch of noise drawn from the aLIGO design PSD. The Wiener-Khinchin theorem states that the ACF and PSD are a Fourier transform pair. We thus compute our ACF by Fourier transforming a noise PSD which is a Welch estimate from 4096 seconds of noise; to allow for acyclic boundary conditions, each periodogram is defined on a noise segment much longer than our analysis segment and we truncate the resulting ACF to match the length of our analysis~\cite{AnalyzingBHRingdowns}.

\subsubsection{Amplitude, frequency, and damping rate posteriors}

First we analyze a no-noise injected signal~\cite{data_release} consisting of a single damped sinusoid with parameters similar to the fundamental prograde ${(\ell,m)=(2,2)}$ QNM of GW150914~\cite{GW150914_detection} (see Table~\ref{tab:injected_params}). The results of this analysis are shown in Fig.~\ref{fig:downsampling_noNoise}. We investigated a range of optimal network SNRs from roughly 10 to 110 (not all shown). We show the highest SNR injection simply because conditioning-induced changes to the posteriors are the most severe at high SNRs. The results for all investigated SNRs are qualitatively similar. We use a prior on the QNM amplitude and phase which is approximately flat around the true values of these parameters, and a flat prior on the black hole mass and spin. The amplitude prior we choose is designed to maximize computational efficiency; its computational expense does not scale with the number of QNMs in the model, whereas the expense of a flat prior scales proportionally with the number of QNMs~\cite{amplitude_prior_note}.

The true parameters of the injected sinusoid can be reasonably recovered when no downsampling is applied. Without downsampling, the posteriors peak at their true values for the frequency and damping rate, and close to the true value for the amplitude (because its prior is not entirely flat). However, downsampling changes the posteriors relative to their ``unconditioned'' forms, and these changes increase in severity with increasing downsampling factor. Furthermore, such posterior changes depend in a nontrivial way on the parameters of the QNMs: for example, as shown in Fig.~\ref{fig:downsampling_noNoise} and App.~\ref{sec:AppendixA}, different QNM phases shift signal power between the signal's high and low frequency tails in the Fourier domain, and signals with phases that maximize the power at low frequencies can be downsampled more aggressively without altering the posteriors. While most of the above cases involve the peaks of the posteriors shifting, we have also observed other cases where downsampling instead causes the posteriors to tighten around their peak values, which is a similarly undesirable behavior.

\subsubsection{Tests of general relativity}
Next, we show that the posterior changes introduced by our current downsampling and filtering methods have ramifications for tests of general relativity. In Fig.~\ref{fig:TGR_downsampling}, we perform a mock test of general relativity on a Kerr-like injection of 2 damped sinusoids, using the TGR model of Eq.~\eqref{eq:TGR_fgamma}. When fitting a signal with multiple QNMs and downsampling, the conditioning-induced changes to the frequency and damping-rate posteriors of each individual QNM can cause a Kerr-consistent spectrum to look erroneously like a beyond-Kerr signal. Fig.~\ref{fig:TGR_downsampling} demonstrates that this downsampling effect can be clearly seen in the TGR posteriors of a single signal. The SNR of 57 in this signal is higher than what is currently accessible, but is lower than what will likely be seen in next-generation detectors~\cite{LIGOvoyager, ETDesignReport, CE_Horizon_study, Baker:2019nia}; at the higher SNRs expected in the future, tests of general relativity using the ringdown will only become more sensitive to such issues. Similarly, in a hierarchical analysis where the posteriors of many events are effectively stacked \cite{Isi_HierarchicalTestGR}, such posterior changes could propagate and become more severe if downsampling is applied too aggressively.

\subsubsection{Filtering comparison}
\begin{figure*}
    \centering
    \includegraphics[width=\textwidth]{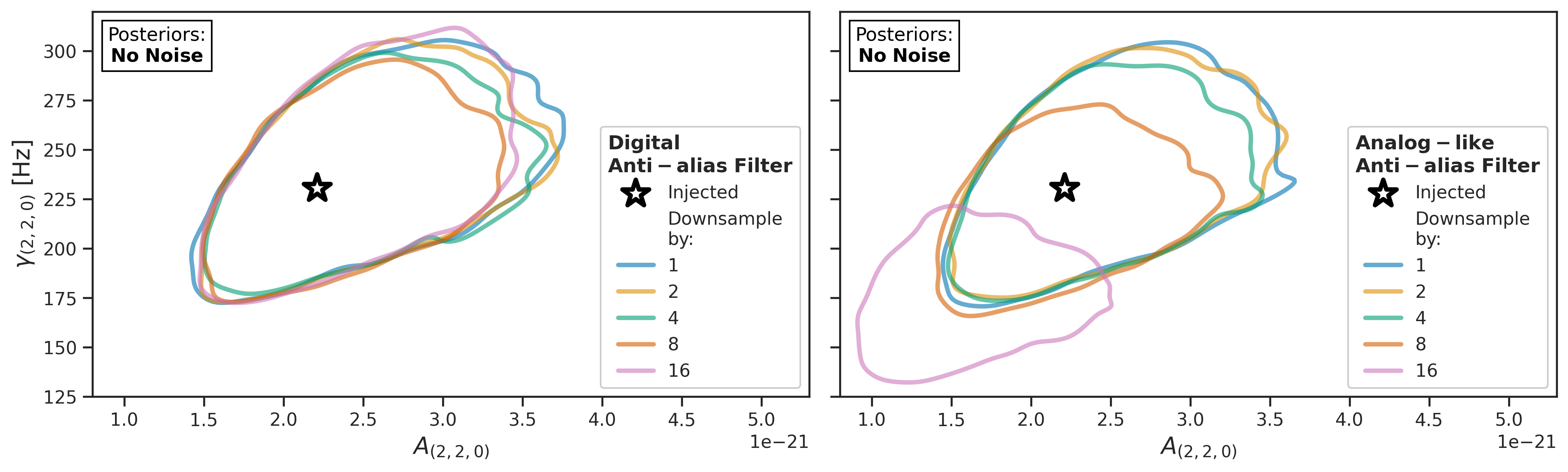}
    \cprotect\caption{Analog-like anti-alias filters, i.e. those with a transient frequency-domain response at their roll-off frequency, induce more significant posterior changes than our ``digital filter", which has an instantaneous response~(as described in Sec.~\ref{sec:DS_Filt}). Here we compare the changes induced by both filters on the amplitude and damping rate posteriors of a GW150914-like $(\ell,m,n) = (2,2,0)$ QNM, in an optimal~network~SNR~$\sim17$ no-noise injection of both the  $(2,2,0)$ and the $(2,2,1)$ (see Table~\ref{tab:injected_params}). 90$\%$ credible contours are shown. The same noise PSD of Fig.~\ref{fig:downsampling_noNoise} is used here to calculate the covariance matrix $\mathbf{C}$ which enters our likelihood, Eq.~\eqref{eq:logL}. ~\textit{Left:}~Downsampling with our digital filter.~\textit{Right:}~Downsampling with an analog-like filter, in this case the Chebyshev filter of the $\textsc{scipy}$~\cite{scipy} package's \verb|signal.decimate| method. This analog-like method is commonly implemented in ringdown analyses. For all downsampling factors, the digital filter retains more posterior support for high amplitudes and high damping rates.}
    \label{fig:digital_vs_analog}
\end{figure*}

Finally, in Fig.~\ref{fig:digital_vs_analog} we compare the effects of downsampling with our digital filter as opposed to more traditional analog-like filters for anti-aliasing. We find generally that analyses with the digital filter have less severe posterior changes when downsampling, which is why the previous results in this section were only shown with the digital filter applied. Although analog-like filters are commonly used in ringdown data analysis,\footnote{Such anti-aliasing filters are also routinely used in more general Fourier-domain gravitational wave analyses; however, for those analyses, the effect of the slow filter response is mitigated by simply truncating the likelihood integration at a frequency much lower than Nyquist.}
there does not appear to be any strong motivation for this choice of filter over one more suitable to the analysis problem, such as our digital filter for example. The digital filter is now the default in our \textsc{ringdown} package~\cite{ringdown_code}. The SNR of the signal in Fig.~\ref{fig:digital_vs_analog} is much lower than that of Fig.~\ref{fig:downsampling_noNoise}, and thus the systematic effects induced by downsampling with the digital filter in the latter are more readily apparent.

\subsubsection{Log-likelihood changes from conditioning}
For a given SNR, it seems that small downsampling factors may introduce acceptably small posterior changes in some cases. As discussed in Sec.~\ref{sec:likelihood}, the posteriors from analyses with conditioning which satisfy Eq.~\eqref{eq:logL_differences} should be considered statistically equivalent to the posteriors of unconditioned analyses. For our no-noise studies, Eq.~\eqref{eq:logL_differences} can be computed by just checking for absolute differences in $\text{SNR}_\text{opt}^2(\mathbf{s})$ before and after conditioning. We save the explicit computation of this quantity in specific injections for future work

\section{Data Segment Duration and PSD lines}\label{sec:PSD_lines}

The LIGO noise PSD contains many powerful and narrow spikes~\cite{guidetoLV_noise, narrowlines_PhysRevD.97.082002}, as shown in Fig.~\ref{fig:PSD_estimation}, which we will refer to as ``lines''. These lines can significantly reduce the total recovered SNR in our analysis if both (1) the lines have frequencies near the central frequencies of QNMs in the signal, and (2) the analysis data segment is not chosen to be long enough to resolve the width of the line in the Fourier domain, or equivalently capture the full length of the whitened signal in the time domain. The longer data segments needed to compensate for the effects of lines increase computational expense. 

In this section, we demonstrate the interplay of PSD lines and data segment duration choices by studying injected signals, and we then discuss methods for managing the SNR loss from lines. Unlike in the case of downsampling and filtering (see Sec.~\ref{sec:injectionstudies_downsample}), the operation of selecting the duration is performed identically in our model and the data. Thus, we only expect the widths of the posteriors in our no-noise studies to depend on the data segment duration; their peaks should remain in the same place regardless of the data segment length.

\begin{figure*}
    \centering
    \includegraphics[width=\textwidth]{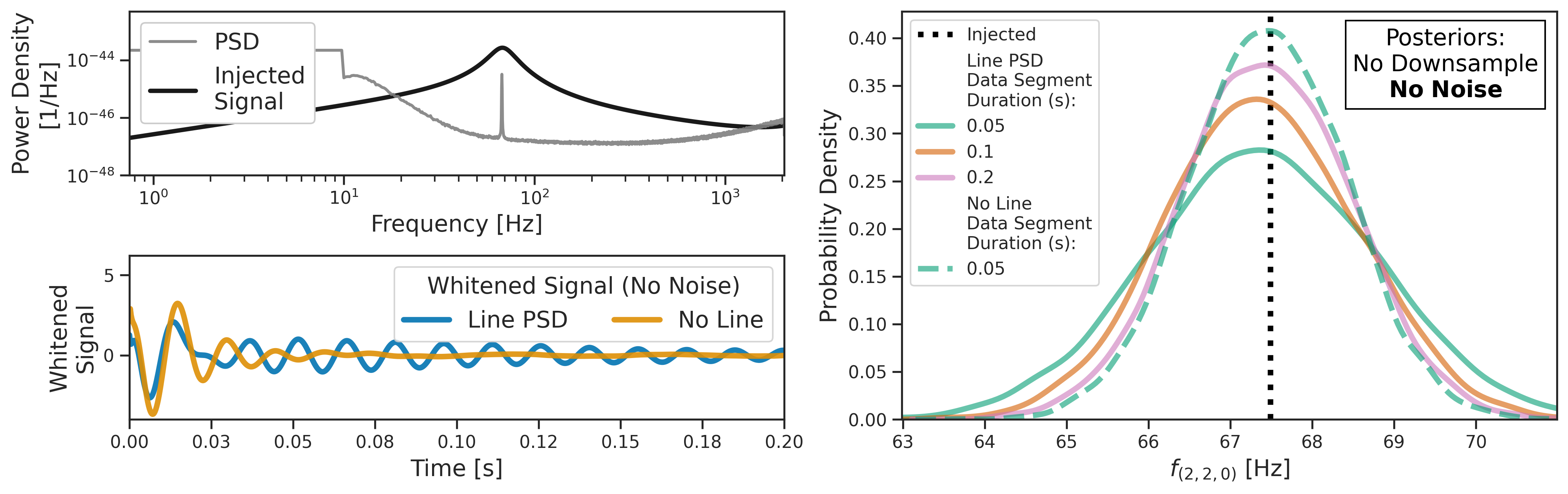}
    \caption{Lines in the noise PSD, when overlapping with central signal frequencies, can reduce recovered SNR if the analyzed data segment is shorter than the whitening timescale associated with the line. \textit{Top~Left:}~We inject a no-noise GW190521-like single damped sinusoid (see Table~\ref{tab:injected_params}), with optimal ${\text{SNR}\sim 18}$ when whitened using a noise covariance matrix derived from a PSD with a Lorentzian line (Eq.~\eqref{eq:lorentzian_line_PSD}) at the signal's central frequency. The PSD is shown in grey, and the signal $f|\tilde{\mathbf{s}}|^2$ in black (see Sec.~\ref{sec:freq_domain}). \textit{Bottom~Left:}~The whitened signal, with and without PSD line. Signal whitened with the line PSD will ring for a longer time. \textit{Right:} With the line PSD, SNR accumulates slowly with increasing data segment length. Correspondingly, the posteriors slowly shrink in width with increasing data segment length. If the line were removed before analysis, much shorter data segments could be used while recovering the full SNR (see dashed line posterior).}
    \label{fig:PSD_line}
\end{figure*}

To begin, it is useful to build intuition in both the time and frequency domains for the interaction of our analysis with PSD lines. In the time domain: lines have a long whitening response time (see Fig.~\ref{fig:PSD_line}), meaning that any signal power near a line's central frequency will be spread over a long timescale when whitened. Note that the whitened data and whitened signal model are the relevant time series in our analysis, since they are the only quantities that enter our likelihood in Eq.~\eqref{eq:logL2}. In order to recover the full SNR, our analysis has to analyze a data segment that is long enough to capture all of the whitened signal ringing in the time domain. Especially if there are lines in the PSD, this data segment may be much longer than the un-whitened signal. In the frequency domain: the frequency bin width equals the inverse of the time domain data segment duration, and thus shorter data segments correspond to larger frequency bin widths. Widening a bin that contains a line will cause the line's power to spread over more frequencies, suppressing the signal in that bin and lowering the recovered SNR.

Lines are especially problematic for longer-lived and lower-frequency signals like GW190521~\cite{GW190521g_DiscoveryPaper, GW190521g_properties}. Longer-lived signals are narrower in the frequency domain, meaning that the PSD line, if aligned with the signal, can potentially cover a larger fraction of the signal power if the frequency bins are made too wide. Lower-frequency signals can also peak closer to the AC mains electricity line~\cite{LIGOScientificNOISE:2019hgc, Davis:2018yrz, Tiwari:2015ofa, PhysRevD.101.042003} (60 Hz in LIGO, 50 Hz in Virgo), which is one of the most powerful narrow-band noise features.

To capture the full SNR when the PSD has lines near the signal, an alternative approach to analyzing longer data segments is to instead clean the lines from the data before analysis. This type of noise subtraction is possible because the lines are highly coherent in the time domain. As an example, a code package to remove individual Lorentzian lines is available here~\cite{LineCleaner_code}. Care must be taken when removing lines, to avoid simultaneous subtraction of signal power. Line subtraction allows for much shorter data segments to be analyzed while also recovering more total SNR than an analysis with the line could achieve.

\begin{figure}
    \centering    \includegraphics[width=0.5\textwidth]{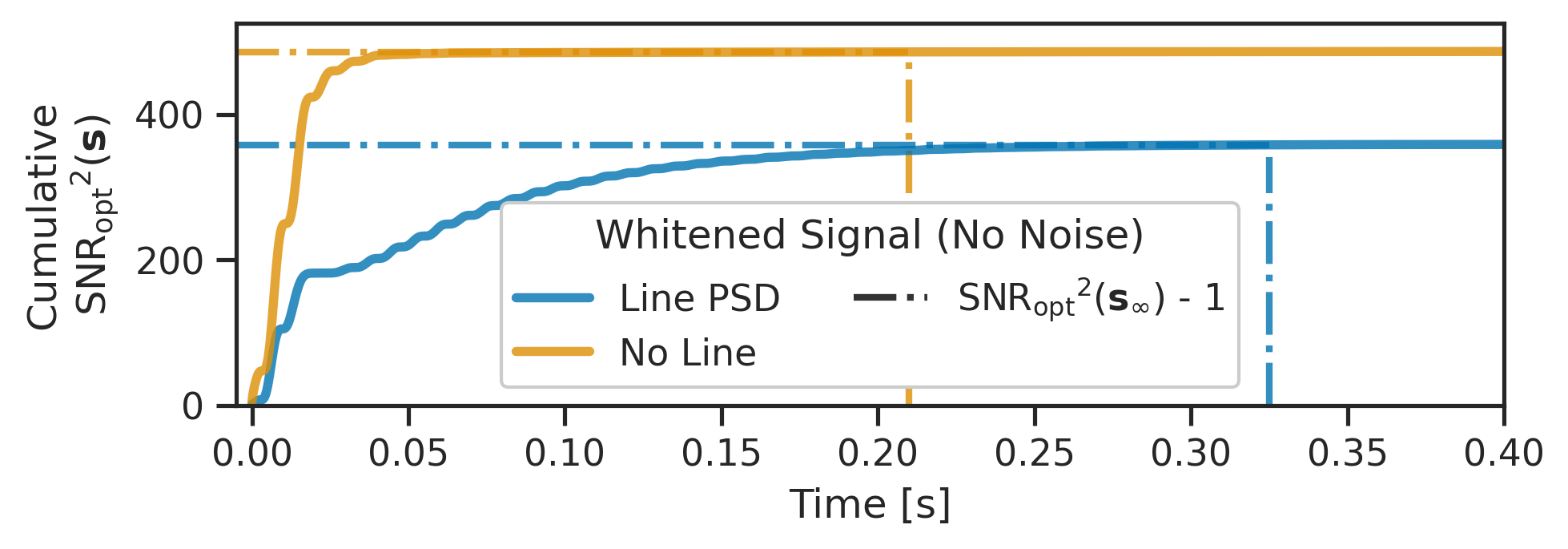}
    \caption{Informed by Eq.~\eqref{eq:logL_differences}, we can estimate the minimum necessary data segment length for our no-noise analysis by computing the amount of time required to accumulate $\text{SNR}^2_\text{opt}-1$ of an effectively infinite-duration data segment, denoted in the legend above by $\mathbf{s}_\infty$. Here we perform such an estimate for the signal in Fig.~\ref{fig:PSD_line}. Our heuristic gives estimates of $\sim 0.2$~s (no line in the PSD) and $\sim 0.325$~s (line) for the data segment length, which is in broad agreement with our findings in Fig.~\ref{fig:PSD_line}, although we note that this heuristic is somewhat conservative. The above figure also shows that if the line in the PSD of Fig.~\ref{fig:PSD_line} were removed before analysis, the total recoverable SNR would increase by $\sim 10\%$.}
    \label{fig:cumulativeSNR_line}
\end{figure}

\subsection{Injection studies: data segment duration}\label{sec:injectionstudies_duration}

To better understand the whitening effects of PSD lines, we compare the analysis of no-noise injections with and without a PSD line. The results of this comparison are shown in Fig.~\ref{fig:PSD_line}. The signal is injected into a single interferometer and at a native sample rate of 4096 Hz (not downsampled) for computational efficiency, but otherwise the analysis is similar to that described in Sec.~\ref{sec:injectionstudies_downsample}.

To compute the ACF in our likelihood~(Eq.~\eqref{eq:logL2}), we use a noise PSD which is a Welch estimate from 4096 seconds of noise drawn from the smooth design PSD for aLIGO~\cite{AdvancedLIGOScientific:2014pky, designPSD_1, designPSD_2} with and without an added Lorentzian line~\footnote{Strictly speaking, Eq.~\eqref{eq:lorentzian_line_PSD} approximates the full Lorentzian function that is derived from the stochastically driven damped harmonic oscillator equation. The approximation in Eq.~\eqref{eq:lorentzian_line_PSD} is valid for frequencies near the central line frequency. At frequencies far from the central frequency, where the approximation becomes worse, the aLIGO design PSD should dominate the Lorentzian.}, i.e.,
\begin{equation}
    \text{PSD}(f) = \text{PSD}_{\substack{\text{LIGO} \\ \text{Design}}}(f)+ \frac{P\Gamma}{2\pi\big[(f-f_0)^2+(\Gamma/2)^2\big]},
    \label{eq:lorentzian_line_PSD}
\end{equation}
where ${\Gamma=0.05}$~Hz, ${P=10^{-45}}$, and $f_0$ is chosen to be the central frequency of the injected signal. For data segment duration $T$, we use $T \in \{0.05~\text{s},0.1~\text{s},0.2~\text{s}\}$. We perform high-pass filtering at a roll-off frequency of $\text{max}(T^{-1}, 10~\text{Hz})$ and we then ``patch" the PSD below the roll-off frequency. We do not perform any downsampling or low-pass filtering.

As shown in Fig.~\ref{fig:PSD_line}, having a line in the PSD that overlaps with the signal can lead to significant SNR loss unless the analysis is performed on a data segment long enough to capture the prolonged ringing of the whitened signal. SNR loss widens the posteriors of any recovered parameters, since posterior width scales as $\sim 1/\text{SNR}$. For $T=0.05$~s with the line in the PSD, the SNR loss is roughly $25\%$. However, when performing an analysis without the line (i.e., assuming the line has been perfectly subtracted), a 0.05 s long data segment recovers more SNR than a 0.2 s long segment whitened with the line. The 0.05 s analysis is also in principle 16 times quicker to run than the 0.2 s analysis, since the number of computations in our analysis scales as $\mathcal{O}(N^2)$ for $N$ data points.

In Fig.~\ref{fig:cumulativeSNR_line}, informed by Eq.~\eqref{eq:logL_differences}, we estimate the minimum necessary data segment length required to recover the full SNR. Such an estimation can be made before performing any analysis. To do this, first choose a reasonable signal model which could approximately represent the true ringdown signal. For example, one might choose the ringdown portion of an IMR fit to the data. Next, whiten this chosen model and (since this is a no-noise injection) calculate its cumulative $\text{SNR}^2_\text{opt}(\mathbf{s})$ as a function of time. The time at which the cumulative  $\text{SNR}^2_\text{opt}(\mathbf{s})$ function becomes ``flat" denotes the data segment duration that is needed to recover the total signal power. More formally, from Eq.~\eqref{eq:logL_differences}, we aim to have a data segment long enough to accumulate at minimum 1 less than the total available $\text{SNR}^2_\text{opt}(\mathbf{s})$. Regarding motivation for this specific criterion: we expect that this criterion should reasonably well bound the largest possible log-likelihood changes at all points in the neighborhood of the parameters of our chosen model to be at most $\mathcal{O}(1)$, and we also expect that such log-likelihood changes should be reasonably uniform throughout the neighborhood for sufficiently high SNR signals. So long as these assumptions are valid, then we expect this criterion should satisfy Eq.~\ref{eq:logL_differences}. An analysis on a data segment of this length should be effectively equivalent to an analysis of an infinite-duration data segment. If there is noise in the data, then the full likelihood must be evaluated as in Eq.~\eqref{eq:logL_differences}, which effectively means one must also consider the accumulation of the matched-filter SNR (see the discussion under Eq.~\eqref{eq:logL2}).

The minimum necessary data segment durations that we estimate in Fig.~\ref{fig:cumulativeSNR_line} closely correspond to those we found empirically in the analysis of Fig.~\ref{fig:PSD_line}, although our heuristic estimates in this case of ${\sim} 0.2$~s (no line) and ${\sim} 0.325$~s (line) are somewhat conservative. Especially when using the no-line PSD, we can go to data segments around half as long as what our estimate suggests and lose only an additional value of 2 in $\text{SNR}^2_\text{opt}(\mathbf{s})$, a fractionally small value. Judging by eye, one might even conclude from the yellow curve in Fig.~\ref{fig:cumulativeSNR_line} that 0.05 seconds is sufficient for this signal when using the no-line PSD; we find correspondingly that longer segments than this produce posteriors that are by eye indistinguishable. Fig.~\ref{fig:cumulativeSNR_line} also demonstrates that removing the line would increase the total recoverable SNR by ${\sim} 10\%$.

\begin{table}[htbp]
    \small 
    \setlength{\tabcolsep}{4pt} 
    \renewcommand{\arraystretch}{1.5} 
    \centering
    \begin{tabular}{|>{\centering}m{0.7cm}|*{4}{c|}}
        \hline
        \multirow{2}{*}{Fig.} & \multicolumn{4}{c|}{Injected Signal} \\
        \cline{2-5}
         & $f$ [Hz] & $\tau$ [ms]& $\phi$ [rad]& $A/10^{-21}$ \\
        \hline
        \ref{fig:downsampling_noNoise},~\ref{fig:appendix_downsampling_noNoise} & 246.7 & 4.3 & 5.4 or 1.2 & 9.1 \\
        \hline
        \ref{fig:TGR_downsampling} & (246.7, 241.7) & (4.3, 1.4) & (5.4, 2.0) & (8.8, 13.8) \\
        \hline    
        \ref{fig:digital_vs_analog} & (246.7, 241.7) & (4.3, 1.4) & (1.2, -1.5) & (2.3, 3.5) \\
        \hline
        \ref{fig:PSD_line}, \ref{fig:cumulativeSNR_line} & 67.5 & 15.5 & 5.4 & 1.2 \\
        \hline
    \end{tabular}
    \caption{Injected parameters of analyzed signals in this paper. The multi-mode injections have $f$ and $\tau$ derived from a single Kerr mass and dimensionless spin using the \textsc{qnm} package~\cite{qnmpackage_Stein}. Numbers in table are rounded.}
    \label{tab:injected_params}
\end{table}

\section{Conclusion}\label{sec:conclusion}

Although downsampling, filtering, and short data segment durations can lead to computational gains in our time-domain ringdown analysis, these operations can undesirably alter our posterior distributions if applied without care. The main takeaways of this paper are: (1) standard downsampling and filtering techniques can not only undesirably alter posteriors, but can even introduce systematic error if not applied carefully; (2) analyzed data segments may need to be much longer than the duration of the unwhitened signal in order to capture the full signal-to-noise ratio, owing to narrow lines in the noise power spectral density; and (3) we offer useful mathematical (Eq.~\eqref{eq:logL_differences}) and graphical diagnostics (Sec.~\ref{sec:freq_domain}, Figs.~\ref{fig:PSD_estimation},~\ref{fig:downsampling_noNoise},~\ref{fig:PSD_line},~\ref{fig:cumulativeSNR_line},~and~\ref{fig:appendix_downsampling_noNoise}) to guide analysts and connect ringdown analysis intuitions with those of more familiar frequency-domain gravitational-wave data analyses.

Currently implemented downsampling and filtering methods in many ringdown data analyses~\cite{Pyring, AnalyzingBHRingdowns, CapanoGW190521g_330, IsiNoHair_GW150914, Carullo:2019flw, Siegel:2023lxl} are designed in such a way that they are not applied identically to the data and the model. We showed in Fig.~\ref{fig:downsampling_noNoise} that, when applied too aggressively, these current methods can not only alter the posteriors by changing the signal-to-noise ratio, but they can also move the peaks of the posteriors even in the absence of noise. In Fig.~\ref{fig:TGR_downsampling} we demonstrated that such undesirable effects could lead to false claims of deviations from general relativity. Adherence to the criterion of Eq.~\eqref{eq:logL_differences} should prevent such erroneous general relativity violations. For the lower signal-to-noise ratios that LIGO signals currently have, such systematic biases should not be of much concern, but as the signals get louder with detector upgrades and next-generation instruments~\cite{LIGOvoyager, ETDesignReport, CE_Horizon_study, Baker:2019nia}, such systematics may become more of an issue. In principle, it should be possible in the future to implement alternative downsampling and filtering methods which treat the data and model in the same way. With such alternative methods, downsampling and filtering would only alter the posteriors when these operations threw out signal information; this would just reduce the signal-to-noise ratio, and would not lead to false general relativity violations.

In Fig.~\ref{fig:digital_vs_analog}, we showed that our ``digital" anti-alias filter method better preserves the structure of our posteriors when compared with other more traditional filtering methods which are typically used. The digital filter simultaneously downsamples and low-pass filters by applying a top-hat in the frequency domain up to Nyquist and then only Fourier transforming frequencies below Nyquist.

We also highlighted in this paper how sharp noise power spectral density lines like those found in LIGO-Virgo-KAGRA data have very long data-whitening timescales in the time domain. Because of this, if these noise lines overlap with central ringdown signal frequencies, our analysis will not recover the full signal unless segments of data much longer than the unwhitened signal are analyzed. This effect is demonstrated in Fig.~\ref{fig:PSD_line}. In Fig.~\ref{fig:cumulativeSNR_line}, we showed how one can estimate the minimum data segment length needed in order to capture all of the signal. The longer data segments needed to capture the full signal-to-noise ratio when a line is present can easily increase the computational expense of our analysis by an order of magnitude compared to when there is no line. However, we note that noise lines can be removed before analysis due to their coherence in the time domain, which allows for short data segments to be used while also increasing the total recoverable SNR significantly.

The data conditioning challenges we have addressed herein have already been relevant in several analyses of actual LIGO-Virgo-KAGRA data~~\cite{Pyring, AnalyzingBHRingdowns, CapanoGW190521g_330, IsiNoHair_GW150914, Carullo:2019flw, Siegel:2023lxl, Haitian1, Haitian2, Haitian3}, and are likely to affect more in the future. Many different time-domain ringdown analyses~\cite{AnalyzingBHRingdowns, Pyring, Bhagwat_RingdownSBITimeDomain}, time-domain-equivalent ringdown analyses~\cite{CapanoGW190521g_330}, and more general truncated time-domain gravitational-wave analyses~\cite{Miller:2023ncs} are susceptible to the sorts of systematics and computational expense issues we have discussed here.

\begin{acknowledgments}
We thank Yuri Levin for inspiring the PSD line subtraction method implemented in~\cite{LineCleaner_code}. We also thank both Yuri Levin and Neil Cornish for helpful comments on early versions of the manuscript, as well as Gregorio Carullo, Yifan Wang, and Collin Capano for stimulating discussion. The Flatiron Institute is a division of the Simons Foundation. H.S. is supported by Yuri Levin's Simons Investigator Award 827103.

\textit{Software:} \textsc{ringdown}~\cite{ringdown_code}, \textsc{arviz}~\cite{arviz_2019}, \textsc{pymc}~\cite{Pymc}, \textsc{numpyro}~\cite{numpyro}, \textsc{pyro}~\cite{pyro}, \textsc{jax}~\cite{jax2018github}, \textsc{seaborn}~\cite{Seaborn}, \textsc{matplotlib}~\cite{matplotlib}, \textsc{jupyter}~\cite{jupyter}, \textsc{numpy}~\cite{numpy}, \textsc{scipy}~\cite{scipy}, \textsc{qnm}~\cite{qnmpackage_Stein},  \textsc{pandas}~\cite{pandas}, \textsc{python3}~\cite{Python3}.

\end{acknowledgments}

\appendix

\begin{figure*}
    \centering
    \includegraphics[width=\textwidth]{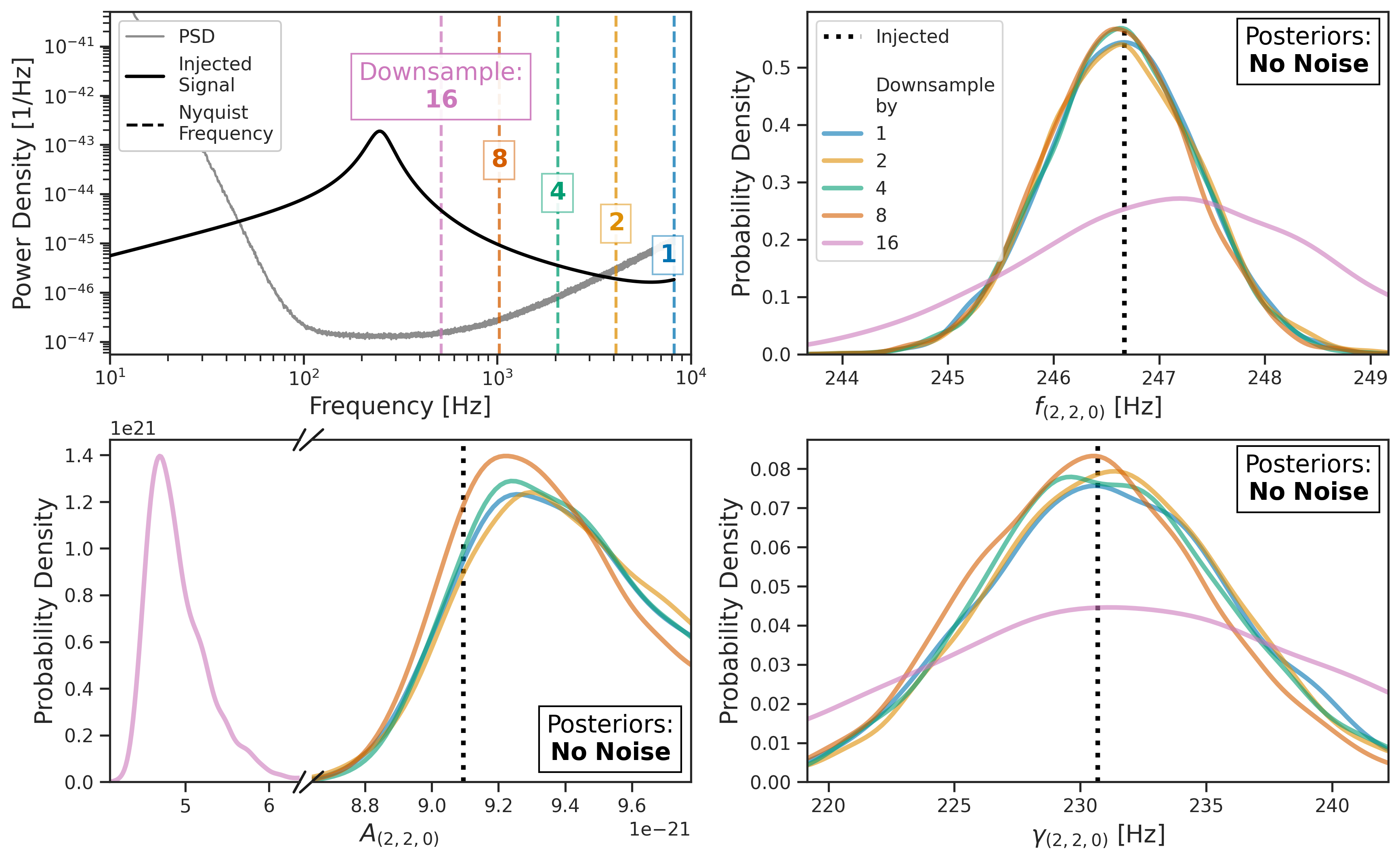}
    \caption{Posterior alteration from downsampling is dependent on the true QNM parameters. Here we perform the same analysis as in Fig.~\ref{fig:downsampling_noNoise}, except that the QNM phase $\phi = 1.2$ rad. The optimal network SNR of this signal is $\sim $100. The posteriors do not change from their unconditioned values until a larger downsampling factor than in Fig.~\ref{fig:downsampling_noNoise}. At top left, the signal is plotted as $f|\tilde{\mathbf{s}}|^2$ (black). In the frequency domain, different $\phi$ values can move the signal power from high to low frequencies (see Sec.~\ref{sec:freq_domain}); compared to Fig.~\ref{fig:downsampling_noNoise}, more of the signal power in this case lies at low frequencies, allowing for more aggressive downsampling without changing the posteriors. The amplitude posterior does not peak exactly at truth when no downsampling is applied; the amplitude prior is not fully flat around the true value.}
    \label{fig:appendix_downsampling_noNoise}
\end{figure*}
\section{\label{sec:AppendixA}Downsampling phase dependence}

In Fig.~\ref{fig:appendix_downsampling_noNoise}, we demonstrate the phase-dependent behavior of single damped sinusoid posteriors when downsampling. We perform the same analysis as in Fig.~\ref{fig:downsampling_noNoise}, except that the QNM phase $\phi = 1.2$ rad. Different phases can distribute signal power between high and low frequencies differently. Given that in this case, more of the signal power lies at lower frequencies, one might naïvely expect that this signal is less sensitive to downsampling when compared with the signal shown in Fig.~\ref{fig:downsampling_noNoise}. This intuition appears to be correct.

This example illustrates the nontrivial and signal-dependent behavior of our analysis when downsampling is applied. We find no definitive rule that would help in predicting the exact behavior of the posteriors when downsampling an arbitrary signal, which is why we encourage conservative downsampling whenever possible. In real signals, it is still not known what phases should be expected for a given set of QNMs, and it is likely that the phases will vary widely depending on the exact details of the binary black hole merger in question, as well as the time in the signal at which the model is fit.

\section{\label{sec:AppendixB}Reduction of number of terms in model}

The model of Eqs.~\ref{eq:h+_hx_def},~\ref{eq:h+_hx_def2},~\ref{eq:signal_model} is a sum of sets of damped sine and cosine functions with the same frequency. Sums of sinusoids with the same frequency can always be re-expressed as a single sinusoid~\cite{mathworld_harmonicadditiontheorem}; we do this below, as it is an instructive exercise. In the following, we drop QNM indices for convenience, and $\omega = 2\pi f$.

First, we re-express the cosine terms in Eq.~\eqref{eq:h+_hx_def} as sine terms, such that
\begin{equation}
    \cos(\omega t + \phi) = -\sin(\omega t + \phi - \pi/2).
\end{equation}
Next, we plug Eq.~\eqref{eq:h+_hx_def} into Eq.~\eqref{eq:signal_model}:
\begin{align}
\begin{split}
    \textbf{s}_I =&A~\exp(-\mathbf{t}/\tau) \cdot \\
    \Big\{\big[&-F^I_+\text{cos}(\theta) - F^I_\times\sin(\theta)\big]\sin(\omega \mathbf{t} + \phi - \pi/2)\\
    &+\epsilon\,\big[-F^I_+\sin(\theta) + F^I_\times\text{cos}(\theta)\big]\sin(\omega \mathbf{t} + \phi) \Big\}.
\end{split}
\end{align}

We then make use of the identity
\begin{align}
\begin{split}
        c~\sin(x+\theta_c) =& a~\sin(x+\theta_a) + b~\sin(x+\theta_b),
\end{split}
\end{align}
where
\begin{align}
\begin{split}
        c^2 =& a^2 + b^2 + 2ab\,\text{cos}(\theta_a - \theta_b),\\
        \text{tan}(\theta_c) =&\frac{a\,\sin(\theta_a) + b\,\sin(\theta_b)}{a\,\cos(\theta_a) + b\,\cos(\theta_b)}.
\end{split}
\end{align}
This gives us a simplified formula (Eq.~\eqref{eq:reduced_signal_model}) for the signal in a single detector, and expresses the 4 observable quantities per QNM in the data (which are an amplitude, damping rate, frequency, and phase) in terms of the 6 parameters per QNM which we aim to constrain in our model (amplitude, damping rate, frequency, phase, ellipticity, and polarization angle).

\begin{widetext}
\begin{eqnarray}
    \textbf{s}_I &=& \sum_j  A^\text{reduced}_j~\exp(-\mathbf{t}/\tau_j)~\sin(\omega_j \mathbf{t} + \phi^\text{reduced}_j) \nonumber \\
    \frac{A^2_\text{reduced}}{A^2} &=& \text{cos}(\theta)^2 (F_+^2+\epsilon^2 F_\times^2)+\sin(\theta)^2(F_\times^2+\epsilon^2 F_+^2)+2F_+F_\times\text{cos}(\theta)\sin(\theta)(1-\epsilon^2) \nonumber \\
    \text{tan}(\phi^\text{reduced}) & = & \frac{-\sin(\phi - \pi/2)~\big[F_+ \text{cos}(\theta) + F_\times\sin(\theta)\big]+ \epsilon~\sin(\phi)~\big[F_\times \text{cos}(\theta) - F_+\sin(\theta)\big]}
    {-\text{cos}(\phi - \pi/2)~\big[F_+ \text{cos}(\theta) + F_\times\sin(\theta)\big]+ \epsilon~\text{cos}(\phi)~\big[F_\times \text{cos}(\theta) - F_+\sin(\theta)\big]}
    \label{eq:reduced_signal_model}
\end{eqnarray}
\end{widetext}
Eq.~\eqref{eq:reduced_signal_model} shows that our most general Kerr model is underconstrained unless measurements are made in at least 2 detectors (i.e. unless there are antenna patterns $F_{+/\times}$ for at least 2 inteferometers). However, if one uses a model where a pair of quantities is assumed to be known, e.g. when analyzing spin-aligned or anti-aligned binaries we might choose to use a fixed viewing angle and linear polarization, a single-detector measurement can then constrain all the parameters of our model.

One might naïvely expect that it is generally preferable to draw posterior samples in terms of the reduced parameters, since for a single interferometer this reduced model has less dimensions to sample from. However, the dimensionality of the sampling space for Eq.~\eqref{eq:reduced_signal_model} grows with the number of interferometers, whereas the dimensionality of Eq.~\eqref{eq:h+_hx_def} does not, meaning that when analyzing data detected in more than 2 interferometers it is more efficient to use Eq.~\eqref{eq:h+_hx_def}.


\clearpage

\bibliography{bibliography}

\end{document}